\PassOptionsToPackage{unicode}{hyperref}
\PassOptionsToPackage{hyphens}{url}
\PassOptionsToPackage{dvipsnames,svgnames,x11names}{xcolor}
\documentclass[
  12pt]{article}

\usepackage{amsmath,amssymb,amsthm}
\usepackage{makecell}
\usepackage{multirow}
\usepackage{iftex}
\ifPDFTeX
  \usepackage[T1]{fontenc}
  \usepackage[utf8]{inputenc}
  \usepackage{textcomp} 
\else 
  \usepackage{unicode-math}
  \defaultfontfeatures{Scale=MatchLowercase}
  \defaultfontfeatures[\rmfamily]{Ligatures=TeX,Scale=1}
\fi
\usepackage{lmodern}
\ifPDFTeX\else  
\fi
\IfFileExists{upquote.sty}{\usepackage{upquote}}{}
\IfFileExists{microtype.sty}{
  \usepackage[]{microtype}
  \UseMicrotypeSet[protrusion]{basicmath} 
}{}
\makeatletter
\@ifundefined{KOMAClassName}{
  \IfFileExists{parskip.sty}{%
    \usepackage{parskip}
  }{
    \setlength{\parindent}{0pt}
    \setlength{\parskip}{6pt plus 2pt minus 1pt}}
}{
  \KOMAoptions{parskip=half}}
\makeatother
\usepackage{xcolor}
\makeatletter
\ifx\paragraph\undefined\else
  \let\oldparagraph\paragraph
  \renewcommand{\paragraph}{
    \@ifstar
      \xxxParagraphStar
      \xxxParagraphNoStar
  }
  \newcommand{\xxxParagraphStar}[1]{\oldparagraph*{#1}\mbox{}}
  \newcommand{\xxxParagraphNoStar}[1]{\oldparagraph{#1}\mbox{}}
\fi
\ifx\subparagraph\undefined\else
  \let\oldsubparagraph\subparagraph
  \renewcommand{\subparagraph}{
    \@ifstar
      \xxxSubParagraphStar
      \xxxSubParagraphNoStar
  }
  \newcommand{\xxxSubParagraphStar}[1]{\oldsubparagraph*{#1}\mbox{}}
  \newcommand{\xxxSubParagraphNoStar}[1]{\oldsubparagraph{#1}\mbox{}}
\fi
\makeatother

\usepackage{longtable,booktabs,array}
\usepackage{calc} 
\usepackage{etoolbox}
\makeatletter
\patchcmd\longtable{\par}{\if@noskipsec\mbox{}\fi\par}{}{}
\makeatother
\IfFileExists{footnotehyper.sty}{\usepackage{footnotehyper}}{\usepackage{footnote}}
\makesavenoteenv{longtable}
\usepackage{graphicx}
\makeatletter
\def\maxwidth{\ifdim\Gin@nat@width>\linewidth\linewidth\else\Gin@nat@width\fi}
\def\maxheight{\ifdim\Gin@nat@height>\textheight\textheight\else\Gin@nat@height\fi}
\makeatother
\setkeys{Gin}{width=\maxwidth,height=\maxheight,keepaspectratio}
\makeatletter
\def\fps@figure{htbp}
\makeatother

\makeatletter
\@ifpackageloaded{caption}{}{\usepackage{caption}}
\AtBeginDocument{%
\ifdefined\contentsname
  \renewcommand*\contentsname{Table of contents}
\else
  \newcommand\contentsname{Table of contents}
\fi
\ifdefined\listfigurename
  \renewcommand*\listfigurename{List of Figures}
\else
  \newcommand\listfigurename{List of Figures}
\fi
\ifdefined\listtablename
  \renewcommand*\listtablename{List of Tables}
\else
  \newcommand\listtablename{List of Tables}
\fi
\ifdefined\figurename
  \renewcommand*\figurename{Figure}
\else
  \newcommand\figurename{Figure}
\fi
\ifdefined\tablename
  \renewcommand*\tablename{Table}
\else
  \newcommand\tablename{Table}
\fi
}
\@ifpackageloaded{float}{}{\usepackage{float}}
\floatstyle{ruled}
\@ifundefined{c@chapter}{\newfloat{codelisting}{h}{lop}}{\newfloat{codelisting}{h}{lop}[chapter]}
\floatname{codelisting}{Listing}

\makeatother
\makeatletter
\@ifpackageloaded{caption}{}{\usepackage{caption}}
\@ifpackageloaded{subcaption}{}{\usepackage{subcaption}}
\makeatother

\ifLuaTeX
  \usepackage{selnolig}  
\fi
\usepackage[]{natbib}
\usepackage{bookmark}
\usepackage{bm}
\IfFileExists{xurl.sty}{\usepackage{xurl}}{} 
\hypersetup{
  pdftitle={Title},
  pdfauthor={Author 1; Author 2},
  pdfkeywords={3 to 6 keywords, that do not appear in the title},
  colorlinks=true,
  linkcolor={blue},
  filecolor={Maroon},
  citecolor={Blue},
  urlcolor={Blue},
  pdfcreator={LaTeX via pandoc}}

\newcommand{\anon}{1}
\newtheorem{theorem}{Theorem}

\newtheorem{proposition}{Proposition}

\begin{document}

\def\spacingset#1{\renewcommand{\baselinestretch}%
{#1}\small\normalsize} \spacingset{1}


\if1\anon
{
  \title{\bf Homogeneity Pursuit of Panel Data with Sparsity}
  \author{Jiming Zhu, Xiong Jin, and Yang Han\thanks{Corresponding author. Email: yang.han@manchester.ac.uk}\hspace{.2cm}\\
    Department of Mathematics, The University of Manchester, UK}
    \date{}
  \maketitle} \fi

\if0\anon
{
  \bigskip
  \bigskip
  \bigskip
  \begin{center}
    {\LARGE\bf Homogeneity Pursuit of Panel Data with Sparsity}
\end{center}
  \medskip
} \fi

\bigskip
\begin{abstract}
In panel data analysis, studying individual attributes is important for both methodological development and practical applications. Homogeneity in individual attributes and sparsity in coefficients are therefore important topics, as they help uncover relationships among individuals and identify the truly relevant covariates. However, many existing approaches do not explicitly exploit the potential grouping information contained in separately fitted individual-level penalized estimators to identify latent groups. To address this gap, we propose a hierarchical group recovery approach based on sparse initial estimation to jointly identify latent groups and sparse parameter structures in panel data. We first introduce a basic version that can efficiently identify homogeneity and sparsity. Based on this basic version, we further introduce a local search step to improve the performance of our method. Compared with existing methods, our approach shows the best overall performance when jointly considering estimation accuracy and computational efficiency. The simulation results demonstrate the effectiveness and advantages of our approach over existing approaches. We further apply the proposed approach to a real dataset to show its practical applicability. 
\end{abstract}

\noindent%
\vfill

\newpage
\spacingset{1.8} 

\section{Introduction}\label{Introduction}
Classical statistical approaches for panel data usually assume that all individuals share the same coefficient vector. The advantage of this assumption is that more data can be used to fit the model, but its limitation is that it ignores possible heterogeneity across individuals. On the other hand, estimating the coefficients separately for each individual is limited by the sample size of each individual and therefore may fail to achieve high estimation accuracy. Sample size limitations are also common in real-world problems. Such limitations may arise not only because the data available for some individuals are inherently limited, but also because some observations may not be suitable for use. For example, in stock price prediction, we usually prefer to use recent data for forecasting, since data from earlier periods may reflect an economic environment that is different from the current one, which may affect the prediction performance. Homogeneity pursuit in panel data, which assumes that there exists a latent group structure among individuals, provides a useful compromise. It can account for heterogeneity across groups while still maintaining a sufficient sample size for parameter estimation within each group. In addition, assuming sparsity in coefficients can not only reduce the number of parameters to be estimated, thereby reducing the sample-size requirement of the model and helping achieve accurate estimation with a smaller sample, but also identify the covariates that are truly relevant to the model.

Homogeneity pursuit is an important topic in panel data analysis. It aims to uncover the latent group structure among individuals that share the same coefficient vector, and thus improve estimation accuracy by increasing the effective sample size. There is a large body of literature on homogeneity pursuit, and we review several related studies below and refer readers to the references therein. \cite{kasahara2009nonparametric} studied non-parametric identification of finite mixture models of dynamic discrete choices. \cite{lin2012estimation}, \cite{bonhomme2015grouped}, and \cite{sarafidis2015partially} applied K-means or K-means-type algorithms to detect homogeneity in individual attributes. \cite{shen2015inference} applied the Structured Logistic-Normal Mixture Model, which combined the finite mixture model with logistic regression, to study subgroups. \cite{bester2016grouped} showed the benefit of grouping individuals under the assumption that the group structure is known. \cite{su2016identifying} proposed the Classifier-Lasso (C-Lasso) to determine the group memberships of individuals of the panel data. \cite{ke2016structure} assumed that the coefficient elements can be partitioned into homogeneous groups, and used a binary segmentation algorithm to detect the underlying cluster structure. \cite{ma2017concave} and \cite{MaHuangZhangLiu+2020} proposed an approach based on a concave pairwise fusion penalty to detect the latent group structure in the intercepts and slopes. \cite{ando2017clustering} proposed to model the panel data with an asset pricing model. This model used unobservable group-specific factors to reflect the group structure. \cite{wang2018homogeneity} extended the CARDS algorithm, which was introduced in \cite{ke2015homogeneity}, to the panel data and the slope coefficient vector, and proposed the Panel-CARDS algorithm to cluster coefficient vectors. Following \cite{ke2016structure}, \cite{wang2021identifying} proposed an approach that applied the sequential binary segmentation algorithm to the nonlinear panel data models to detect the homogeneity structure.  \cite{xiao2021homogeneity} made the same assumption of homogeneity in coefficient elements as \cite{ke2016structure} and studied the homogeneity structure in large-scale panel data with interactive effects and heavy-tailed errors. \cite{lu2021multiply} proposed an approach that integrates median regression with the concave fusion penalty and multiply robust weights to detect intercept homogeneity in panel data with dropouts. \cite{wang2024panel} focused on the problem that the group membership and the corresponding coefficients changed at a change time point and developed an algorithm to estimate parameters and the change point. \cite{MEHRABANI20231464} used a pairwise fusion penalty, specifically an adaptive group fused Lasso penalty, to detect the latent group structure in slopes.  \cite{Han03042026} proposed an approach, RBIN, to detect homogeneity in short panel data.

Sparsity in coefficients has been studied for many years. Classical methods include \cite{tibshirani1996regression} (Lasso), \cite{fan2001variable} (SCAD), and \cite{zhang2010nearly} (MCP). The joint consideration of homogeneity in attributes and sparsity in coefficients in panel data has started to receive attention in recent years; see \cite{huang2023integrative} and \cite{wang2024homogeneity}, among others, and the references therein. However, to the best of our knowledge, existing methods often make little use of the potential grouping information contained in penalized estimators obtained separately for each individual to identify latent groups. Although applying penalized estimation to each individual may lead to low estimation accuracy and inaccurate sparsity identification because of the limited sample size for each individual, the resulting estimators may still contain useful information about the group structure. Such information can help us identify homogeneity more efficiently. In practice, this means that the computational time can be greatly reduced while maintaining high estimation accuracy. 

In this paper, we develop a new hierarchical group recovery framework based on sparse initial estimation to jointly identify latent homogeneous groups and sparse parameter structures in panel data. When carrying out penalized estimation for each individual, we apply the marginal correlation ranking approach of Sure Independence Screening in \cite{fan2008sure}. This approach is particularly useful for adjusting the order when updating parameters in the Coordinate Descent (CD) algorithm, so as to reduce the bias of the penalized estimation due to the update order effect. We propose two versions of the new framework, the Basic Stepwise Penalized Regression and Grouping (BSPRG) method and the more advanced Stepwise Penalized Regression and Grouping with Local Search (SPRG-LS) method. For SPRG-LS, we further add a local search step to correct possible clustering errors in the group structure obtained from the initial estimators. In simulation studies, SPRG-LS performs better than existing methods in most settings, with higher Normalized Mutual Information (NMI) values and Estimation Error (EE) values closer to those of the oracle estimator. It also has a substantial advantage in terms of computational cost.

The remainder of the paper is organized as follows: In \hyperref[Models]{Section \ref*{Models}}, we introduce the models used in this paper. In \hyperref[Estimation Procedure]{Section \ref*{Estimation Procedure}}, we give a detailed description of our proposed approaches. In \hyperref[Asymptotic Properties]{Section \ref*{Asymptotic Properties}}, we present the asymptotic properties of our proposed approaches. In \hyperref[Simulation Study]{Section \ref*{Simulation Study}}, we present some simulation results. In \hyperref[Real Data Analysis]{Section \ref*{Real Data Analysis}}, we show the results of applying our approach to real data. In \hyperref[Conclusion]{Section \ref*{Conclusion}}, we draw conclusions. The Supplementary Material includes the algorithm description, additional simulation results, theoretical analysis, and additional information on real data analysis.
\section{Models}\label{Models}
For individual \(i\) at time point \(t\), let \(Y_{it}\) be the response and \(\bm{X}_{it}\) be the covariate vector, \(i = 1,\ldots,N, t = 1,\ldots,T\). We assume that the model is given as 
\begin{align}
    Y_{it} = \bm{X}_{it}^{\top}\bm{\beta}_{i} + \epsilon_{it},\label{model formula 1}
\end{align}
where \(\epsilon_{it}\) is the error term. Model (\ref{model formula 1}) can be expressed in the following matrix form
\begin{align}
    \bm{Y}_i = \bm{X}_i\bm{\beta}_i + \bm{\epsilon}_{i},\, i = 1,\ldots, N, \label{Basic model}
\end{align}
where \(\bm{Y}_i = (Y_{i1},\ldots,Y_{iT})^{\top}\), \(\bm{X}_i = (\bm{X}_{i1},\ldots,\bm{X}_{iT})^{\top}\), and \(\bm{\epsilon}_i = (\epsilon_{i1},\ldots,\epsilon_{iT})^{\top}\). We assume that there exists a latent group structure in \(\{\bm{\beta}_1,\ldots, \bm{\beta}_N\}\), given by
\begin{align}
    \bm{\beta}_i = 
    \begin{cases}
        \bm{\alpha}_{G_1}, \text{if } i \in G_1, \\
        \bm{\alpha}_{G_2}, \text{if } i \in G_2, \\
        \ldots\\
        \bm{\alpha}_{G_{K_0}}, \text{if } i \in G_{K_0},\\
    \end{cases}
\end{align}
where \(i=1,\ldots,N\), \(\{\bm{\alpha}_{G_1},\ldots,\bm{\alpha}_{G_{K_0}}\}\) denotes the list of cluster-specific coefficient vectors corresponding to the cluster list \(\{G_1,\ldots,G_{K_0}\}\), and \(K_0\) is the number of clusters.

Also, we assume that each cluster-specific coefficient vector \(\bm{\alpha}_{G_k}\) is sparse, \(k = 1,\ldots,K_0\). We let \(\operatorname{supp}(\bm{\alpha}_{G_k})\) denote the support of \(\bm{\alpha}_{G_k}\), i.e., the set of indices corresponding to the non-zero entries of \(\bm{\alpha}_{G_k}\), and \(s_{G_k} = \vert\operatorname{supp}(\bm{\alpha}_{G_k})\vert\) is the number of indices in \(\operatorname{supp}(\bm{\alpha}_{G_k})\). \(\bm{\alpha}_{G_k}^{(\mathrm{I})} = (\bm{\alpha}_{G_k,j})_{j \in \operatorname{supp}(\bm{\alpha}_{G_k})} \in \mathbb{R}^{s_{G_k}}\) denotes the subvector of \(\bm{\alpha}_{G_k}\) indexed by \(\operatorname{supp}(\bm{\alpha}_{G_k})\), and \(\bm{\alpha}_{G_k}^{(\mathrm{II})} = (\bm{\alpha}_{G_k,j})_{j \in \operatorname{supp}(\bm{\alpha}_{G_k})^{c}} = \bm{0}_{p-s_{G_k}}\) denotes the subvector of \(\bm{\alpha}_{G_k}\) indexed by the complement of \(\operatorname{supp} (\bm{\alpha}_{G_k})\). Similarly, for individual \(i \in G_k, k \in \{1,\ldots,K_0\}\), we denote \(\bm{\beta}_{i}^{(\mathrm{I})} = \bm{\alpha}_{G_k}^{(\mathrm{I})} \in \mathbb{R}^{s_{G_k}}\) as  the subvector of \(\bm{\beta}_{i}\) indexed by \(\operatorname{supp}(\bm{\beta}_{i})\), and \(\bm{\beta}_{i}^{(\mathrm{II})} = \bm{\alpha}_{G_k}^{(\mathrm{II})} = \bm{0}_{p - s_{G_k}}\) as the subvector of \(\bm{\beta}_{i}\) indexed by the complement of \(\operatorname{supp} (\bm{\beta}_{i})\).
\section{Estimation Procedure}\label{Estimation Procedure}
\subsection{Basic Stepwise Penalized Regression and Grouping (BSPRG)}\label{BSPRG}
\subsubsection{Initial Estimation}\label{Initial Estimation}
For the linear regression model with \(T \geq p\), the estimator of the \(i\)th individual's coefficient vector is
\begin{align}
    \hat{\bm{\beta}_i} &= \operatorname*{arg\,min}_{\bm{\beta} \in \mathbb{R}^{p}} (2T)^{-1}\sum_{t = 1}^T(Y_{it} - \bm{X}_{it}^{\top}\bm{\beta})^2\nonumber \\
    &= (\bm{X}_i^{\top}\bm{X}_i)^{-1}\bm{X}_i^{\top}\bm{Y}_i ,\quad i = 1,\ldots,N.\nonumber
\end{align}
However, note that under our assumptions, \(T\) may not be greater than \(p\). When \(T < p\), the estimation above does not work. Combining this with the potential sparsity in \(\bm{\beta}_i\), applying penalized regression to obtain initial estimators is a natural choice, and the Smoothly Clipped Absolute Deviation (SCAD) penalty proposed by \cite{fan2001variable} is widely used because of its good performance in addressing sparsity. After adding the SCAD penalty, the initial estimator \(\hat{\bm{\beta}}_i^{(0)}\) is obtained by solving
\begin{align}
   \hat{\bm{\beta}}_i^{(0)} = \operatorname*{arg\,min}_{\bm{\beta} \in \mathbb{R}^{p}} \frac{1}{2T}\sum_{t = 1}^T(Y_{it} - \bm{X}_{it}^{\top}\bm{\beta})^2 + \sum_{j=1}^{p}p_{\lambda_i}(\vert\beta_{j}\vert),\quad i = 1,\ldots,N, \label{Initial estimation}
\end{align}
where \(p_{\lambda_i}(\cdot)\) is the SCAD penalty.

The choice of the tuning parameter \(\lambda_i\) is a key part of the practical implementation of penalized regression, because it is closely related to the accuracy of the coefficient estimators. Given the features of our setting, we use the BIC criterion proposed by \cite{fan2013tuning} to select \(\lambda_i\). Suppose that there are \(N_{\lambda}\) candidate values of \(\lambda_i\), denoted by \(\bm{\Lambda}=\{\lambda^1,\ldots,\lambda^{N_{\lambda}}\}\). Let \(\hat{\bm{\beta}}_i^{(0),\lambda_i}\) be the initial estimator of \(\bm{\beta}_i\) obtained by solving the optimization problem (\ref{Initial estimation}) under \(\lambda_i\). For model (\ref{Basic model}), the BIC value for each individual under \(\lambda_i\) is given by
\begin{align}
    BIC_i^{\lambda_i} = \log (\frac{\|\bm{Y}_i - \bm{X}_i\hat{\bm{\beta}}_i^{(0),\lambda_i}\|_{2}^2}{T}) + \frac{a_Ts^{\lambda_i}_i}{T},\quad i = 1,\ldots,N, \quad \lambda_i \in \bm{\Lambda}, \label{BIC_lambda}
\end{align}
where \(a_T\) is a diverging sequence with respect to \(T\), and \(s^{\lambda_i}_i\) is the number of non-zero parameters in \(\hat{\bm{\beta}}_i^{(0),\lambda_i}\). In our approach, we set \(a_{T} = \log(\log(T))\log(p)\), following \cite{fan2013tuning}. We then select \(\lambda_i\) by minimizing \(BIC_i^{\lambda_i}\).

Next, we discuss how to solve the optimization problem (\ref{Initial estimation}). For its computational implementation, a variety of solvers can be used, such as the \texttt{ncvreg} package (\cite{breheny2011coordinate}) in \textsf{R}. To the best of our knowledge, the Coordinate Descent (CD) algorithm is widely used in many solvers, including \texttt{ncvreg}, to obtain the solution of the penalized regression problem. For details of the CD algorithm, see \cite{breheny2011coordinate}. When the sample size is large, the CD algorithm performs well. However, when the sample size is limited and the support of \(\bm{\beta}_i = (\beta_{i1},\ldots,\beta_{ip})^{\top}\), denoted by \(\operatorname{supp}(\bm{\beta}_i)=\{j:\beta_{ij}\neq 0, 1 \leq j \leq p\}\), contains indices with relatively large values, especially those close to \(p\), the order in which the coefficients are updated can substantially affect the estimation accuracy of the CD algorithm. We observe this phenomenon for the initial estimators in some simulation studies.

To solve the above problem, we propose to use the marginal correlation ranking approach of Sure Independence Screening (SIS) proposed by \cite{fan2008sure} to adjust the covariate update order in coordinate descent. For each \(v \in \{1,\ldots,p\}\), we calculate the Pearson correlation coefficient \(\rho_{iv}\) between \((\bm{X}_i)_{\cdot v}\) and \(\bm{Y}_i\), where $(\bm{X}_i)_{\cdot v}$ denotes the \(v\)th column of \(\bm{X}_i\), and obtain \(\{\rho_{i1},\ldots,\rho_{ip}\}\). Let \((j_1,\ldots,j_p)\) be the permutation of \((1,\ldots,p)\) such that \(\vert\rho_{ij_1}\vert \geq \vert\rho_{ij_2}\vert \geq \ldots \geq \vert\rho_{ij_p}\vert\). We then reorder the columns of \(\bm{X}_i\) according to this permutation and obtain \(\mathring{\bm{X}}_i=((\bm{X}_{i})_{\cdot j_1},\ldots,(\bm{X}_{i})_{\cdot j_p}\)). Based on \(\mathring{\bm{X}}_i\), we apply the CD algorithm to solve the optimization problem and obtain the estimator \(\hat{\mathring{\bm{\beta}}}^{(0)}_i=(\hat{\beta}^{(0)}_{ij_1},\ldots,\hat{\beta}^{(0)}_{ij_p})^\top\). Since \(\hat{\mathring{\bm{\beta}}}^{(0)}_i\) is defined under the reordered covariates, we map it back to the original order of the covariates to get \(\hat{\bm{\beta}}^{(0)}_i\).
\subsubsection{Clustering and Re-estimation of Coefficients}
Based on the initial estimators \(\{\hat{\bm{\beta}}_1^{(0)},\ldots,\hat{\bm{\beta}}_N^{(0)}\}\), we proceed to perform clustering. We apply the agglomerative hierarchical clustering with complete linkage and the \(\ell_{\infty}\)-distance. Denote the candidate numbers of clusters as \(\mathcal{K} = \{K_{\min},\ldots,K_{\max}\}\) that we search through in our approach. For any \(K \in \mathcal{K}\), we apply the agglomerative hierarchical clustering to \(\{\hat{\bm{\beta}}_1^{(0)},\ldots,\hat{\bm{\beta}}_N^{(0)}\}\) and obtain the estimated partition \(\hat{\bm{\Delta}}_K = \{\hat{G}_1,\ldots,\hat{G}_{K}\}\). For \(k \in \{1,\ldots,K\}\), we estimate the cluster-specific coefficient vector by
\begin{align}
    \hat{\bm{\alpha}}_{\hat{G}_k} = \operatorname*{arg\,min}_{\bm{\alpha} \in \mathbb{R}^p}\frac{1}{2T\vert\hat{G}_k\vert}\sum_{i \in \hat{G}_k}\|\bm{Y}_i - \bm{X}_{i}\bm{\alpha}\|_2^2 + \sum_{j=1}^{p}p_{\lambda_{\hat{G}_k}}(\vert\alpha_{j}\vert),\label{Alpha update}
\end{align}
where \(\vert\hat{G}_k\vert\) is the number of individuals in \(\hat{G}_k\).

To select the tuning parameter \(\lambda_{\hat{G}_k}\) used in the optimization problem (\ref{Alpha update}), we use the BIC criterion defined by formula (\ref{BIC_lambda}). When estimating \(\bm{\alpha}_{G_k}\), we still adjust the covariate update order in the CD algorithm, as described in \hyperref[Initial Estimation]{Section \ref*{Initial Estimation}}.

Next, we need to determine a proper number of clusters. Here, following \cite{wang2024homogeneity}, we use the \(mBIC\) criterion to select the number of clusters. For each \(K \in \mathcal{K}\) and its corresponding partition \(\hat{\bm{\Delta}}_K\), the \(mBIC\) value is defined as
\begin{align}
    mBIC_{\hat{\bm{\Delta}}_K } = \log (\frac{\sum_{k=1}^K\sum_{i \in \hat{G}_k}\|\bm{Y}_{i} - \bm{X}_{i}\hat{\bm{\alpha}}_{\hat{G}_k}\|_2^2}{NT}) + \frac{\log(\log(NT + p))\log(NT)\sum_{k=1}^Ks_{\hat{G}_k}}{NT}\label{mBIC_cluster},
\end{align}
where \(s_{\hat{G}_k}\) is the number of non-zero parameters in \(\hat{\bm{\alpha}}_{\hat{G}_k}\). We obtain the estimated number of clusters by minimizing \(mBIC_{\hat{\bm{\Delta}}_K}\). In \hyperref[Asymptotic Properties]{Section \ref*{Asymptotic Properties}}, we prove that, under appropriate assumptions, as \(N,T\to\infty\), the probability that the \(mBIC\) criterion selects the true number of clusters tends to 1.
\subsubsection{Further Clustering}
With sufficiently large \(N\) and \(T\), accurate estimation of the latent group structure and the coefficients can be obtained through the above steps. However, in applications, due to the limited accuracy of the initial estimators \(\{\hat{\bm{\beta}}^{(0)}_1,\ldots,\hat{\bm{\beta}}^{(0)}_N\}\) caused by insufficient sample sizes, it may happen that, for some \(k\in\{1,\ldots, K_0\}\), the true group \(G_k\) is split into subgroups \(\hat{G}_{k_1}\), \(\hat{G}_{k_2}\), or even more subgroups. Nevertheless, \(\hat{G}_{k_1}\) and \(\hat{G}_{k_2}\) may still contain enough individuals, so that the corresponding estimators \(\hat{\bm{\alpha}}_{\hat{G}_{k_1}}\) and \(\hat{\bm{\alpha}}_{\hat{G}_{k_2}}\) are already close to \(\bm{\alpha}_{G_k}\). In this case, the estimated partition is not fully accurate, but the estimators of the cluster-specific coefficient vectors, \(\hat{\bm{\alpha}}_{\hat{G}_{k_1}}\) and \(\hat{\bm{\alpha}}_{\hat{G}_{k_2}}\), are accurate. The reason why \(mBIC\) does not select a partition with fewer clusters that merge \(\hat{G}_{k_1}\) and \(\hat{G}_{k_2}\) is related to the characteristics of agglomerative hierarchical clustering and the accuracy of our initial estimators. For a given number of clusters, agglomerative hierarchical clustering starts with each individual as a separate cluster and, at each step, merges the two clusters with the smallest distance until the target number of clusters is reached. In our approach, we use complete linkage, where the distance between two clusters is defined as the largest distance between individuals in the two clusters. Due to the limited accuracy of the initial estimators \(\{\hat{\bm{\beta}}^{(0)}_1,\ldots,\hat{\bm{\beta}}^{(0)}_N\}\), there may exist individuals in \(\hat{G}_{k_1}\) and \(\hat{G}_{k_2}\) that are far apart. As a result, the complete linkage distance between \(\hat{G}_{k_1}\) and \(\hat{G}_{k_2}\) can be larger than the distance between two estimated clusters belonging to different true clusters. Therefore, when the number of clusters is reduced, agglomerative hierarchical clustering may merge other clusters with smaller distances instead of merging \(\hat{G}_{k_1}\) and \(\hat{G}_{k_2}\), even though they come from the same true cluster. Since \(mBIC\) selects among the partitions generated along this hierarchical path, it may fail to select a partition that correctly merges these subgroups.

Inspired by \cite{wang2018homogeneity}, which includes an additional hierarchical clustering in the final step, we also add a further clustering step. Suppose that, after clustering and re-estimation, we obtain the partition and the corresponding estimated parameters \(\hat{\bm{\Delta}}_K = \{\hat{G}_1,\ldots,\hat{G}_K\}\) and \(\{\hat{\bm{\alpha}}_{\hat{G}_1},\ldots,\hat{\bm{\alpha}}_{\hat{G}_K}\}\). For \(K^{*}\in\{K_{\min},\ldots,K\}\), we perform hierarchical clustering again based on \(\{\hat{\bm{\alpha}}_{\hat{G}_1},\ldots,\hat{\bm{\alpha}}_{\hat{G}_K}\}\), obtaining the resulting partition \(\hat{\bm{\Delta}}_{K^{*}}=\{\hat{G}^{*}_1,\ldots,\hat{G}^{*}_{K^{*}}\}\) and the corresponding cluster-specific coefficient vectors \(\{\hat{\bm{\alpha}}^{*}_{\hat{G}^{*}_1},\ldots,\hat{\bm{\alpha}}^{*}_{\hat{G}^{*}_{K^{*}}}\}\). We compute \(mBIC_{\hat{\bm{\Delta}}_{K^{*}}}\) by (\ref{mBIC_cluster}), and select the \(K^{*}\), together with its corresponding partition and cluster-specific coefficient vectors, that gives the lowest \(mBIC_{\hat{\bm{\Delta}}_{K^{*}}}\). Compared with the previous clustering step, this further clustering step can improve performance because it is based on the more accurate parameter estimators, \(\{\hat{\bm{\alpha}}_{\hat{G}_1},\ldots,\hat{\bm{\alpha}}_{\hat{G}_K}\}\), obtained from the previous clustering step, rather than the initial estimators, \(\{\hat{\bm{\beta}}^{(0)}_1,\ldots,\hat{\bm{\beta}}^{(0)}_{N}\}\), used in the previous clustering step.

To more clearly demonstrate the entire process of BSPRG, we summarize our approach in Algorithm 1 of the Supplementary Material.
\subsection{Stepwise Penalized Regression and Grouping with Local Search (SPRG-LS)}
In \hyperref[BSPRG]{Section \ref*{BSPRG}}, one limitation of BSPRG is the absence of an explicit mechanism for correcting mis-clustered individuals. Inspired by \cite{huang2023integrative}, we improve BSPRG with a local search step to enhance estimation accuracy, which we call SPRG-LS. We keep the initial estimation step, the coefficient re-estimation step, and the further clustering step the same as those in BSPRG, and the main improvement is included in the clustering step.
\subsubsection{Clustering Step of SPRG-LS}
In the clustering step of BSPRG, we only use hierarchical clustering to cluster individuals. Although we apply a further clustering step later, its role is to merge small clusters, rather than to move possibly mis-clustered individuals between clusters.

To provide this missing correction step, we introduce a local search procedure. Suppose that the partition obtained from the clustering step is \(\hat{\bm{\Delta}}_K=\{\hat{G}_1,\ldots,\hat{G}_K\}\), with the corresponding value \(mBIC_{\hat{\bm{\Delta}}_K}\). For \(k\in\{1,\ldots,K\}\) and individual \(i\in\hat{G}_k\), we consider moving \(i\) to another cluster \(\hat{G}_{k'}\), where \(k'\neq k\), and obtain a new partition \(\hat{\bm{\Delta}}'_{K}\). We then re-estimate the cluster-specific coefficient vectors and compute the corresponding value \(mBIC_{\hat{\bm{\Delta}}'_{K}}\). If \(mBIC_{\hat{\bm{\Delta}}'_{K}}<mBIC_{\hat{\bm{\Delta}}_K}\), we accept this movement. We go through all possible choices of \(k'\) and select the one that gives the largest decrease in \(mBIC\) as the final cluster to which the individual \(i\) is moved. We repeat this procedure until the clustering structure no longer changes.

A natural question is which individuals should be considered for reassignment. An option is to check all individuals, as in \cite{huang2023integrative}. However, in our simulation studies, we find that adjusting all individuals increases the computational cost, while the additional gain from adjusting many well-clustered individuals is limited. This motivates us to adjust only those individuals that are more likely to be mis-clustered. We describe the selection rule for these individuals below. First, we consider individuals whose estimators cannot be effectively used. That is, for a partition \(\hat{\bm{\Delta}}_K=\{\hat{G}_1,\ldots,\hat{G}_K\}\) and a cluster \(\hat{G}_k\), \(k\in\{1,\ldots,K\}\), if its corresponding cluster-specific coefficient vector satisfies \(\hat{\bm{\alpha}}_{\hat{G}_k}=\bm{0}\), then this estimated coefficient vector is not appropriate for representing the cluster. This situation may be caused by an inappropriate choice of \(\lambda_{\hat{G}_k}\), which can occur when the sample size within \(\hat{G}_k\) is limited. Therefore, all individuals in such a cluster are included as candidates for adjustment. Second, for clusters with non-zero estimated cluster-specific coefficient vectors, we aim to select the individuals that are more likely to be mis-clustered. For any cluster \(\hat{G}_k\), \(k\in\{1,\ldots,K\}\), we compute the mean squared error (MSE) of each individual \(i\in\hat{G}_k\) as 
\begin{align}
    MSE_{i,\hat{G}_k} = \frac{1}{T}\sum_{t=1}^{T} (Y_{it} - \bm{X}_{it}^{\top}\hat{\bm{\alpha}}_{\hat{G}_k})^2.
\end{align}
For \(i\in\hat{G}_k\), a larger value of \(MSE_{i,\hat{G}_k}\) indicates that this individual is less likely to belong to this cluster. Therefore, we rank the individuals in \(\hat{G}_k\) according to \(MSE_{i,\hat{G}_k}\), which provides an order of their likelihood of being mis-clustered.

A subsequent issue concerns the number of individuals that should be selected. We define a data-driven ratio \(R_K\). The MSE of the partition \(\hat{\bm{\Delta}}_K = \{\hat{G}_1,\ldots,\hat{G}_{K}\}\) can be calculated as
\begin{align}
    MSE_{\hat{\bm{\Delta}}_K} = \frac{1}{NT}\sum_{k=1}^K\sum_{i\in \hat{G}_k}\|\bm{Y}_{i} - \bm{X}_{i}\hat{\bm{\alpha}}_{\hat{G}_k}\|_2^2.
\end{align}
The MSE of the pooled model, where all individuals are in the same cluster and share the same coefficient vector \(\hat{\bm{\alpha}}_{pooled}\), can be calculated as
\begin{align}
    MSE_{pooled} = \frac{1}{NT}\sum_{i=1}^N\|\bm{Y}_i - \bm{X}_i\hat{\bm{\alpha}}_{pooled}\|_2^2.
\end{align}
Then, \(R_K\) can be computed by
\begin{align}
    R_K = r_1 \frac{MSE_{\hat{\bm{\Delta}}_K}}{MSE_{pooled}} \frac{p}{T+p}, \label{R_k formula}
\end{align}
where \(r_1 \in (0,1]\). For cluster \(\hat{G}_k\), we rank the individuals according to \(MSE_{i,\hat{G}_k}\) and select the \(\lceil R_K \vert\hat{G}_k\vert\rceil\) individuals with the highest values as potential candidates for adjustment, where \(\lceil\cdot\rceil\) denotes rounding up to the nearest integer. We consider two aspects when constructing this data-driven ratio: MSE and the relationship between \(T\) and \(p\). The MSE of the pooled model usually serves as an upper benchmark for the MSE we can obtain for \(\{(\bm{X}_i,\bm{Y}_i)\}_{i=1,\ldots,N}\). For a given partition \(\hat{\bm{\Delta}}_{K}\), the ratio of its corresponding MSE to the MSE of the pooled model reflects the improvement of the partition over the pooled model. A lower value indicates that, compared with assigning all individuals into one cluster, the partition can more effectively reflect the latent heterogeneity in the coefficient vectors and reduce the corresponding MSE. Therefore, fewer individuals in such a partition need to be adjusted. The other aspect is the relationship between \(T\) and \(p\), which directly affects the accuracy of \(\{\hat{\bm{\beta}}_1^{(0)},\ldots,\hat{\bm{\beta}}_N^{(0)}\}\). In general, for a fixed \(p\), as \(T\) increases, the value of \(\frac{p}{T+p}\) decreases, and \(\{\hat{\bm{\beta}}_1^{(0)},\ldots,\hat{\bm{\beta}}_N^{(0)}\}\) become more accurate. Therefore, the number of individuals that need to be adjusted is reduced.

After determining the potential individuals that need to be adjusted, we adjust these individuals according to the local search procedure described above, and return the adjusted partition and the corresponding cluster-specific coefficient vectors. Note that, as in \hyperref[BSPRG]{Section \ref*{BSPRG}}, we still apply a further clustering step to the partition obtained after local search, in order to merge clusters that can be combined. The algorithmic structure of the local search procedure is provided in Algorithm 2 of the Supplementary Material.
\section{Asymptotic Properties}\label{Asymptotic Properties}
\subsection{Notation}\label{Notation}
Let 
\[
\bm{\mathcal{G}}^{\circ} = \{G \subseteq \{1,\ldots,N\}: G \neq \emptyset,\ G \subseteq G_k \text{ for some } k\in\{1,\ldots,K_0\}\},
\]
where \(K_0\) is the true number of clusters. For \(G \in \bm{\mathcal{G}}^{\circ}\), the corresponding model is \(\bm{Y}_{G} = \bm{X}_{G}\bm{\alpha}_{G} + \bm{\epsilon}_{G}\), where \(\bm{X}_{G}, \bm{Y}_{G}\) denote the design matrix and the response vector of \(G\), respectively, \( \bm{\alpha}_{G}\) is the true coefficient vector of \(G\), and \(\bm{\epsilon}_{G}\) is the error vector. Let \(\operatorname{supp}(\bm{\alpha}_{G})\) denote the set of indices corresponding to the non-zero entries of \(\bm{\alpha}_{G}\) and let \(s_{G} = \vert\operatorname{supp}(\bm{\alpha}_{G})\vert\) be the number of indices in \(\operatorname{supp}(\bm{\alpha}_{G})\). \(\bm{\alpha}_{G}^{(\mathrm{I})} = (\bm{\alpha}_{G,j})_{j \in \operatorname{supp}(\bm{\alpha}_{G})} \in \mathbb{R}^{s_{G}}\) denotes the subvector of \(\bm{\alpha}_{G}\) indexed by \(\operatorname{supp}(\bm{\alpha}_{G})\), and \(\bm{\alpha}_{G}^{(\mathrm{II})} = (\bm{\alpha}_{G,j})_{j \in \operatorname{supp}(\bm{\alpha}_{G})^{c}} = \bm{0}_{p-s_{G}}\) denotes the subvector indexed by the complement of \(\operatorname{supp} (\bm{\alpha}_{G})\). Note that, when \(G \subseteq G_k\) for a unique \(k \in \{1,\ldots,K_0\}\), we have \(\bm{\alpha}_{G} = \bm{\alpha}_{G_k}\). Accordingly, let \(\bm{X}_{G}^{(\mathrm{I})}\) denote the submatrix of \(\bm{X}_{G}\) formed by the columns indexed by \(\operatorname{supp}(\bm{\alpha}_{G})\), and let \(\bm{X}_{G}^{(\mathrm{II})}\) denote the submatrix of \(\bm{X}_{G}\) formed by the columns indexed by the complement of  \(\operatorname{supp} (\bm{\alpha}_{G})\). When we estimate the coefficient vector based on \(\left(\bm{X}_{G}, \bm{Y}_{G}\right)\), we denote the corresponding tuning parameter by \(\lambda_{G}\), and the corresponding estimator is \(\hat{\bm{\alpha}}_{G}\). \(\hat{\bm{\alpha}}_{G}^{(\mathrm{I})}\) denotes the subvector of \(\hat{\bm{\alpha}}_{G}\) indexed by \(\operatorname{supp} (\bm{\alpha}_{G})\), and \(\hat{\bm{\alpha}}_{G}^{(\mathrm{II})}\) denotes the subvector indexed by the complement of \(\operatorname{supp} (\bm{\alpha}_{G})\). When \(\vert G \vert = 1\), the cluster \(G\) reduces to an individual. In this case, we write
\( \bm{X}_i=\bm{X}_{\{i\}}\in\mathbb{R}^{T\times p}\), \(\bm{Y}_i=\bm{Y}_{\{i\}}\in\mathbb{R}^{T}\), \( \bm{\beta}_i = \bm{\alpha}_{\{i\}}\), \(i = 1,\ldots,N\), and we write the corresponding tuning parameter \(\lambda_i\) in place of \(\lambda_{\{i\}}\) for notational simplicity. For \(k = 1,\ldots,K_0\), the oracle estimator \(\hat{\bm{\alpha}}_{G_k}^{o}\) for the true cluster \(G_{k} \in \bm{\Delta}_{K_0}\) is defined by
\[
    \hat{\bm{\alpha}}_{G_k}^{o,(\mathrm{I})} = \underset{\bm{\alpha}^{(\mathrm{I})} \in \mathbb{R}^{s_{G_k}}}{\arg\min}\frac{1}{2\vert G_k \vert T}\sum_{i \in G_k}\left\|\bm{Y}_i - \bm{X}_{i}^{(\mathrm{I})}\bm{\alpha}^{(\mathrm{I})}\right\|_2^2, \qquad \hat{\bm{\alpha}}_{G_k}^{o,(\mathrm{II})} = \bm{0}.
\]
Let \(\hat{\bm{\Delta}}_{\hat{K}},\hat{K},\{\hat{\bm{\alpha}}_{\hat{G}_1},\ldots,\hat{\bm{\alpha}}_{\hat{G}_{\hat{K}}}\}\) denote the estimated partition, the estimated number of clusters, and the estimated cluster-specific coefficient vectors, respectively. For \(k = 1,\ldots,\hat{K}\), \(s_{\hat{G}_k} = \vert\operatorname{supp}(\hat{\bm{\alpha}}_{\hat{G}_k})\vert\) denotes the number of indices in \(\operatorname{supp}(\hat{\bm{\alpha}}_{\hat{G}_k})\). Without loss of generality, we assume that all columns of the design matrices involved in the section are standardized.

For a matrix \(\bm{B}\), we denote by \(\bm{B}_{i\cdot}\) and \(\bm{B}_{\cdot j}\) the \(i\)th row and the \(j\)th column of \(\bm{B}\), respectively. If \(\bm{B}\) is symmetric, we denote by \(\lambda_{\min}(\bm{B})\) and \(\lambda_{\max}(\bm{B})\) the smallest and largest eigenvalues of \(\bm{B}\), respectively. We define the spectral norm of \(\bm{B}\) by \(\|\bm{B}\|_{2}\), and the Frobenius norm of \(\bm{B}\) by \(\|\bm{B}\|_{F}\). For a matrix \(\bm B=(b_{ij})\in\mathbb R^{n\times p}\), its matrix \(\ell_\infty\)-norm is defined as \(\|\bm{B}\|_\infty = \max\limits_{1\leq i\leq n}\sum_{j=1}^p \vert b_{ij}\vert\). For a vector \(\bm{v} \in \mathbb{R}^q\), we define 
\[
    \|\bm v\|_1=\sum_{j=1}^q \vert v_j\vert,\qquad
    \|\bm v\|_2=\left(\sum_{j=1}^q v_j^2\right)^{1/2},
    \qquad
    \|\bm v\|_\infty=\max_{1\le j\le q}\vert v_j\vert.
\]

Define \(d_{\min} = \min\limits_{k, k' \in \{1,\ldots,K_0\}, k \neq k'}\|\bm{\alpha}_{G_k} - \bm{\alpha}_{G_{k'}}\|_{\infty}\) as the minimal distance between the cluster-specific coefficient vectors \(\{\bm{\alpha}_{G_1},\ldots,\bm{\alpha}_{G_{K_0}}\}\). For \(G \in \bm{\mathcal{G}}^{\circ}\), define \(b_{G,0} = \min\limits_{j \in \operatorname{supp}(\bm{\alpha}_{G})}\vert\alpha_{G,j}\vert\). For the true partition \(\bm{\Delta}_{K_0} = \{G_1,\ldots,G_{K_0}\}\), define \(b_{\min,0} = \min\limits_{k=1,\ldots,K_0}\min\limits_{j \in \operatorname{supp}(\bm{\alpha}_{G_k})}\vert\alpha_{G_k,j}\vert\) as the minimal signal among all non-zero coefficients across the true clusters, and define \(b_{\max,0} = \max\limits_{k=1,\ldots,K_0}\max\limits_{j \in \operatorname{supp}(\bm{\alpha}_{G_k})}\vert\alpha_{G_k,j}\vert\) as the maximal signal among all non-zero coefficients across the true clusters. If \(G \subseteq G_k\), then \(\bm{\alpha}_G = \bm{\alpha}_{G_k}\), and hence \(b_{\min,0} \leq b_{G_k,0} = b_{G,0} \leq b_{\max,0}\).

For the penalty function \(p_{\lambda}(x)\), we denote \(\rho(x,\lambda) = \frac{p_{\lambda}(x)}{\lambda}\). For \(\bm{b} \in \mathbb{R}^{q}\) with \(\|\bm{b}\|_0 = q\), define
\[
   \kappa(\rho;\bm{b}) = \lim_{\epsilon \to 0+}\max_{1\leq j\leq q}\sup_{\substack{t_1,t_2\in(\vert b_j\vert-\epsilon,\vert b_j\vert+\epsilon),t_1<t_2}}-\frac{\rho'(t_2, \lambda) - \rho'(t_1, \lambda)}{t_2 - t_1} .
\]

When both \(A_n\) and \(a_n\) are positive deterministic sequences, we write \(A_n \gg a_n\) if \(\lim\limits_{n \to \infty}a_n/A_n = 0\).
\subsection{Assumptions}\label{Assumption}
\textbf{Assumption 1: } Assume that the candidate number of clusters is searched over a finite set \( \mathcal{K}=\{K_{\min},\ldots,K_{\max}\}\), where \(1 \leq K_{\min} < K_{0} < K_{\max}<\infty\) is independent of \(N\), \(T\), and \(p\). 

We assume that \(\epsilon_{it}\)'s are independent mean-zero random variables with \(\operatorname{Var}(\epsilon_{it}) = \sigma^2\). Furthermore, \(\bm{\epsilon} = (\bm{\epsilon}_{1}^{\top},\ldots,\bm{\epsilon}_{N}^{\top})^{\top}\) has sub-Gaussian tails in the sense that there exists a constant \(a_2 > 0\) such that for any \(\bm{f} \in \mathbb{R}^{NT}\) and \(x > 0\),
\begin{align}
    P(\left\vert\bm{f}^{\top}\bm{\epsilon}\right\vert > \sigma\|\bm{f}\|_2x) \leq 2\exp(-a_2x^2).
\end{align}

For \(s_{G_k}\), \(p\), \(N\), \(T\), we assume that
\begin{align}
    &p = O\left(\left(NT\right)^{a_1}\right)\, \text{ and }\, p \leq \vert G_{\min}\vert T, \label{a_1 there}\\
    &N^{K_{\max} - K_{0}} = o\left(p^{a_2c-1}\right) \, \text{ and }\, \log N = o(T),\\
    &1 \leq s_{G_k} = O(1) < p, \quad \forall k \in \{1,\ldots,K_0\},\\
    &0 < \pi_{\min} \leq \frac{\vert G_{\min}\vert}{N} \leq \frac{\vert G_{\max}\vert}{N} \leq \pi_{\max} < 1,\\
    &b_{\min,0} \gg T^{-\frac{1}{2}}\sqrt{\log p},\\
    &b_{\max,0} \leq a_3,
\end{align}
where \(a_1 \in (0,1)\), and \(c > 0\) is chosen such that \(a_2c >1\). \(\vert G_k\vert\) denotes the number of individuals in the \(k\)th true cluster, \(\vert G_{\min}\vert=\min\limits_{1\leq k\leq K_0}\vert G_k\vert\), \(\vert G_{\max}\vert=\max\limits_{1\leq k\leq K_0}\vert G_k \vert\), and \(\pi_{\min},\pi_{\max} \in (0,1)\). \(a_3 > 0\) is a constant. Since we assume that the latent group structure in \(\{\bm{\alpha}_{G_1},\ldots,\bm{\alpha}_{G_{K_0}}\}\) is distinguishable, the minimal separation \(d_{\min} > 0\) is assumed to be bounded away from \(0\) and independent of \(N\), \(T\), and \(p\).

For any candidate set \(A\subset\{1,\ldots,N\}\) on which the SCAD-penalized regression is performed, let \( n_A=\vert A\vert T\). We assume that the corresponding tuning parameter \(\lambda_{A}\) satisfies, as \(n_A\to\infty\),
\begin{align}
    \lambda_{A}\to0,\qquad
    \lambda_{A}\sqrt{n_A}\to\infty,\qquad
    \lambda_{A}n_A^{\frac{a_1}{2}}\to 0, 
\end{align}
where \(a_1\) is the same as in (\ref{a_1 there}).

\textbf{Assumption 2: } Let \(\bm{\mathcal{G}}_m^{\circ} = \{G \in \bm{\mathcal{G}}^\circ:\ \vert G\vert =m\},\ 1 \leq m \leq \vert G_{\max}\vert\). For any \(G \in  \bm{\mathcal{G}}_m^{\circ}\), we assume that for the corresponding \(\bm{X}_G,\bm{Y}_G,\bm{\epsilon}_G\), the following conditions hold with constants depending only on \(mT\) and the common constants in the assumptions, not on the membership of \(G\):
\begin{align}
    &\frac{1}{mT} \left(\bm{X}_{G,\cdot j}\right)^{\top} \left(\bm{X}_{G,\cdot j}\right) \leq M_1, \qquad j=1,\ldots,p, \label{assump2_colnorm_m}\\
    &\left\|\left\{\left(\bm{X}_G^{(\mathrm{I})}\right)^{\top} \bm{X}_G^{(\mathrm{I})}\right\}^{-1}\right\|_{\infty} \leq C_{1,G} \leq C_{1,m}, \label{assump2_active_inverse_m}\\
    &\left\|\left\{\left(\bm{X}_G^{(\mathrm{II})}\right)^{\top}\bm{X}_G^{(\mathrm I)}\right\}\left\{\left(\bm{X}_G^{(\mathrm{I})}\right)^{\top}\bm{X}_G^{(\mathrm{I})} \right\}^{-1}\right\|_{\infty} \leq C_{2,G} \leq C_{2,m}, \label{assump2_cross_m}\\
    &\frac{1}{mT}\|\bm Y_G\|_2^2=O_p(1). \label{assump2_y_m}
\end{align}
Here \(M_1\in(0,\infty)\), \(C_{1,m} = \sup\limits_{G \in \bm{\mathcal{G}}_{m}^{\circ}}C_{1,G} = O\left(\left(mT\right)^{-1}\right)\), and \(C_{2,m} = \sup\limits_{G \in \bm{\mathcal{G}}_{m}^{\circ}}C_{2,G}\) satisfies \(C_{2,m} \in \left[0,\frac{C}{\rho'(c_0 b_{\min,0})}\right]\) for some constants \(C,c_0\in(0,1)\). The function \(\rho(\cdot)\) is based on the corresponding lower tuning level defined in Assumption 3. In addition, the corresponding tuning parameter \(\lambda_G\) satisfies \(b_{\min,0} \gg \lambda_G\).

Furthermore, for any \(G \in \bm{\mathcal{G}}_m^{\circ}\) with \(mT \geq p\), we assume that
\begin{align}
    0< \frac{1}{a-1} \leq \lambda_{\min}\left(\frac{\bm{X}_G^{\top}\bm{X}_G}{mT}\right) \leq \lambda_{\max}\left(\frac{\bm X_G^{\top}\bm X_G}{mT}\right)\leq M_2,\label{assump2_full_eigen_m}
\end{align}
where \(a>2\) is the fixed SCAD shape parameter, and \(M_2 > 0\). For such \(G\) with \(mT \geq p\), we also assume that
\begin{align}
    (mT)^{(1-c_1)/2} b_{\min,0} \ge M_3 \qquad \text{for some }c_1\in(0,1), \label{assump2_bmin_m}
\end{align}
where \(M_3>0\). Since each \(G \in \bm{\mathcal{G}}_m^{\circ}\) is contained in some true cluster \(G_k\), we have \(s_G = s_{G_k} = O(1)\). Thus, for any fixed \(c_3\in(0,c_1)\),
\begin{align}
    s_G=O((mT)^{c_3}), \qquad G \in \bm{\mathcal{G}}_m^{\circ}.
\end{align}
The tuning parameter \(\lambda_G\), for \(G \in \bm{\mathcal{G}}_m^\circ\), is further assumed to satisfy
\begin{align}
    &\lambda_G = o\left(\left(mT\right)^{-\{1-(c_1-c_3)\}/2}\right), \label{assump2_lambda_rate1_m}\\
    &\frac{p}{(\lambda_G\sqrt{mT})^{2l}} \to 0\label{assump2_lambda_rate2_m}
\end{align}
for some integer \(l>0\).

\textbf{Assumption 3: } Unless otherwise specified, \(\rho(\cdot)\) used below is based on \(\underline{\lambda}_{G}\) defined in this assumption. Let \(u_m \in (0,\infty)\) satisfy \(\lim\limits_{mT \to \infty}u_m = \infty\). For any \(G \in \bm{\mathcal{G}}_m^\circ\), define
\begin{align}
    D_{1,G} = \max\limits_{j\in\{1,\ldots,s_G\}} \|\bm X_{G,\cdot j}^{(\mathrm I)}\|_2, \qquad D_{2,G} = \max\limits_{j\in\{1,\ldots,p-s_G\}} \|\bm X_{G,\cdot j}^{(\mathrm {II})}\|_2.
\end{align}
Additionally, we assume that the sample-size-wise upper bounds satisfy 
\begin{align}
    D_{1,m} = \sup_{G \in \bm{\mathcal{G}}_m^\circ}D_{1,G} = O(\sqrt{mT}), \qquad D_{2,m} = \sup_{G \in \bm{\mathcal{G}}_m^\circ}D_{2,G} = O(\sqrt{mT}).
\end{align}
Assume that 
\begin{align}
   \underline{\lambda}_G = \frac{1}{mT} \frac{(C_{2,G}D_{1,G}+D_{2,G})u_m\sigma} {1-C_{2,G}\rho'(c_0b_{G,0})} \leq \overline{\lambda}_G = \frac{ C_{1,G}^{-1}(1-c_0)b_{G,0} - u_mD_{1,G}\sigma}{mT\rho'(c_0b_{G,0},\overline{\lambda}_G)}.\label{assump3_m_lambdaupper}
\end{align}
Furthermore, we assume that
\begin{align}
   u_m \leq \left\{ \kappa_{m,0}(C_{2,m}D_{1,m}+D_{2,m}) \right\}^{-1}\underline{\phi}_m(1-C)\sigma^{-1},\label{assump3_m_curvature}
\end{align}
where \(\underline{\phi}_m = \inf_{G \in \bm{\mathcal{G}}_m^\circ} \lambda_{\min} \left((\bm{X}_G^{(\mathrm{I})})^\top \bm{X}_G^{(\mathrm{I})}\right)\), \(\kappa_{m,0} = \sup\limits_{G\in\bm{\mathcal{G}}_{m}^{\circ}} \max\left\{\kappa(\rho;\bm{b}): \|\bm{b}-\bm{\alpha}_G^{(\mathrm{I})}\|_\infty \leq (1-c_0)b_{G,0}\right\}\), and \(C,c_0\) are the same constants as in Assumption 2.

Assumptions 2 and 3 are formulated in line with the regularity conditions used to establish the weak oracle property in \cite{lv2009AUNIFIEDAPPROACH} and \cite{fan2010selective}, as well as those used for the oracle property and global optimality of SCAD-penalized estimators in high-dimensional settings in \cite{kim2008smoothly} and \cite{fan2011nonconcave}. These assumptions ensure that there exist initial SCAD estimators satisfying the weak oracle property with a sample-size-wise error bound. Furthermore, the consistent recovery of the latent clustering structure, together with Assumptions 1-3, ensures the existence of cluster-specific SCAD estimators that coincide with the corresponding oracle estimators with probability tending to 1.
\subsection{Main Results}\label{Main Results}
\begin{theorem}\label{Theorem 1}
    Suppose that Assumptions 1-3 are satisfied. For any cluster \(G \in \mathcal{G}_m^{\circ}\), there exists a SCAD-penalized estimator \(\hat{\bm\alpha}_{G}\) with corresponding tuning parameter \( \lambda_G=\underline{\lambda}_G\) and \( u_m^2=c\log p\), where \(a_2c > 1\) and \(a_2\) is the sub-Gaussian tail constant in Assumption 1. Then, as \(mT \to \infty\),
    \[
        \hat{\bm\alpha}_G^{(\mathrm{II})}=\bm0
    \]
    with probability tending to \(1\), and
    \[
        \|\hat{\bm\alpha}_{G}-\bm\alpha_G\|_\infty
        =
        O_p\left((mT)^{-1/2}\sqrt{\log p}\right),
    \]
    where the \(O_p(\cdot)\) bound is sample-size-wise uniform: the deterministic envelope for the rate depends on \(G\) only through its size \(m\) and the constants in Assumptions 1-3, and not through its membership.
\end{theorem}
\hyperref[Theorem 1]{Theorem \ref*{Theorem 1}} establishes a sample-size-wise weak oracle property, following \cite{lv2009AUNIFIEDAPPROACH} and \cite{fan2010selective}, for the SCAD estimator on any \(G \in \mathcal{G}_m^{\circ}\). The result shows that, with probability approaching \(1\), the inactive coefficients are estimated as zero, and the estimation error of the active coefficient vector converges at the rate \((mT)^{-1/2}\sqrt{\log p}\), where \(m\) is the number of individuals in the cluster. The bound is independent of the specific membership of the cluster and depends on the cluster only through its sample size.
\begin{proposition}\label{Proposition 1}
    Suppose that the conditions of \hyperref[Theorem 1]{Theorem \ref*{Theorem 1}} hold. Then, for each \(i=1,\ldots,N\), there exists a SCAD-penalized estimator \(\hat{\bm\beta}^{(0)}_i\), with tuning parameter \(\lambda_i=\underline{\lambda}_i\) and \(u_1^2 = c\log p\), where \(a_2c > 1\), such that, as \(T \to \infty\)
    \[
        \|\hat{\bm{\beta}}^{(0)}_i-\bm{\beta}_i\|_{\infty} = O_p\left(T^{-1/2}\sqrt{\log p}\right).
    \]
    More precisely, there exists a constant \(C_{\beta} > 0\), independent of \(i\), such that 
    \[
        P\left(\|\hat{\bm{\beta}}^{(0)}_{i} - \bm{\beta}_{i}\|_{\infty} \leq C_{\beta}T^{-\frac{1}{2}}\sqrt{\log p}\right) \geq 1 - 2p^{1-a_2c}, \ i = 1,\ldots,N.
    \]
    Moreover, as \(N, T \to \infty\),
    \[
        P\left(\max_{1\leq i\leq N} \|\hat{\bm{\beta}}^{(0)}_i-\bm{\beta}_i\|_{\infty} \leq C_{\beta}T^{-\frac{1}{2}}\sqrt{\log p}\right) \to 1,
    \]
    and hence
    \[
        \max_{1\leq i\leq N} \|\hat{\bm{\beta}}^{(0)}_i-\bm{\beta}_i\|_{\infty} = O_p\left(T^{-1/2}\sqrt{\log p}\right).
    \]
\end{proposition}
\hyperref[Proposition 1]{Proposition \ref*{Proposition 1}} specializes \hyperref[Theorem 1]{Theorem \ref*{Theorem 1}} to singleton clusters. It gives an \(\ell_\infty\) error bound for the initial individual-level SCAD estimators and, by a union bound over individuals, yields a uniform maximum error bound over \(i=1,\ldots,N\). The following theorems are based on estimators satisfying \hyperref[Theorem 1]{Theorem \ref*{Theorem 1}} and \hyperref[Proposition 1]{Proposition \ref*{Proposition 1}}.
\begin{proposition}\label{Proposition 2}
    Suppose that the conditions of \hyperref[Proposition 1]{Proposition \ref*{Proposition 1}} hold. Let \(C_{\beta}\) be the constant appearing in \hyperref[Proposition 1]{Proposition \ref*{Proposition 1}}. Define 
    \[
        \mathcal{E}_{\beta} = \left\{\max_{1\leq i\leq N}\|\hat{\bm{\beta}}^{(0)}_i-\bm{\beta}_i\|_{\infty} \leq C_{\beta}T^{-1/2}\sqrt{\log p}\right\}.
    \]
    Apply the agglomerative hierarchical clustering algorithm with complete linkage to \(\{\hat{\bm{\beta}}_i^{(0)}\}_{i=1}^N\) using the \(\ell_\infty\)-distance. If \(d_{\min}>4C_{\beta}T^{-1/2}\sqrt{\log p}\), then, on the event \(\mathcal{E}_\beta\), when the algorithm is stopped at \(K_0\) clusters, it recovers the true partition up to label permutation:
    \[
        \hat{\bm{\Delta}}_{K_0}=\bm{\Delta}_{K_0}.
    \]
    Consequently, as \(N, T \to \infty\), we have
    \[
        P\left(\hat{\bm{\Delta}}_{K_0}=\bm{\Delta}_{K_0}\right)\to 1.
    \]
\end{proposition}
By \hyperref[Proposition 2]{Proposition \ref*{Proposition 2}}, when the true number of clusters \(K_0\) is given, the agglomerative hierarchical clustering algorithm recovers the true partition with probability tending to \(1\). Therefore, in the following theorem, it suffices to study the SCAD-penalized estimator based on each true cluster \(G_k\).
\begin{theorem} \label{Theorem 2}
    Suppose that Assumptions 1 and 2 hold. For each true cluster \(G_k\), \(k=1,\ldots,K_0\), let \(\hat{\bm{\alpha}}_{G_k}\) denote the SCAD-penalized estimator based on \(G_k\), and let \(\hat{\bm{\alpha}}_{G_k}^{o}\) denote the oracle estimator based on \(G_k\). Then, as \(N, T \to \infty\),
    \[
        P\left(\hat{\bm{\alpha}}_{G_k} = \hat{\bm{\alpha}}_{G_k}^{o}, \ k=1,\ldots,K_0\right) \to 1.
    \]
\end{theorem}
\hyperref[Theorem 2]{Theorem \ref*{Theorem 2}} establishes the oracle property for the SCAD estimator on the true clusters. It shows that, when the true partition is used, the SCAD estimator for each true cluster is identical to the oracle estimator with probability approaching \(1\).

\hyperref[Proposition 2]{Proposition \ref*{Proposition 2}} and \hyperref[Theorem 2]{Theorem \ref*{Theorem 2}} show that, when the true number of clusters is given, we can recover the true latent structure and obtain the oracle estimator for each cluster with probability tending to \(1\). Therefore, whether the proposed \(mBIC\) can select \(K_0\) is important. The following theorem guarantees that \(mBIC\) selects the true number of clusters with probability tending to \(1\).
\begin{theorem}\label{Theorem 3}
    Suppose that Assumptions 1-3 are satisfied. Assume that the candidate number of clusters is searched over a finite set \( \mathcal{K}=\{K_{\min},\ldots,K_{\max}\}\), where \(1 \leq K_{\min} < K_0 < K_{\max}<\infty\) is independent of \(N\), \(T\), and \(p\). For convenience, we define \(mBIC_K := mBIC_{\hat{\bm{\Delta}}_{K}}\). Then, as \(N,T\to\infty\),
    \[
        P(\hat K_{mBIC}=K_0)\to 1,
    \]
    where
    \( \hat K_{mBIC} = \mathop{\arg\min}\limits_{K \in \mathcal K} mBIC_K\).
\end{theorem}
\hyperref[Theorem 3]{Theorem \ref*{Theorem 3}} shows that with probability approaching \(1\), the proposed \(mBIC\) criterion consistently selects the true number of latent clusters.
\section{Simulation Study}\label{Simulation Study}
In this section, we conduct simulation studies to show the effectiveness of our proposed methods. We compare our approaches with tDPLS proposed by \cite{wang2024homogeneity}, and K-SCAD proposed by \cite{huang2023integrative} in Example 1 and Example 2.
\subsection{Example Designs}
\textbf{Example 1: }We set the number of true clusters as \(K_0 = 3\), and the corresponding coefficient vectors are
\[
    \bm{\alpha}_{G_1} = (-0.5,1.5,\bm{0}_{p-2})^{\top}_{p}, \bm{\alpha}_{G_2} = (0.5,0.5,\bm{0}_{p-2})^{\top}_{p}, \text{and }\bm{\alpha}_{G_3} = (1.5,-0.5,\bm{0}_{p-2})^{\top}_{p}.
\]
The proportions of individuals in the three groups are \((0.3,0.3,0.4)\). For the design matrix \(\bm{X}_{it}\), we generate values from the standard Normal distribution \(N(0,1)\). The error term \(\epsilon_{it}\) is generated from a Normal distribution \(N(0,\sigma^2)\). Then the response \(Y_{it}\) can be computed by
\begin{align}
    Y_{it} = \bm{X}_{it}^{\top}\bm{\alpha}_{G_k} + \epsilon_{it}, \quad \text{if \(i \in G_k\)},\nonumber
\end{align}
where \(i = 1,\ldots,N\), \(t = 1,\ldots,T\), \(G_k\) is the true \(k\)th cluster, \(k = 1,2,3\).

We set \(p\) to be \(10\) or \(20\), \(N\) to be \(30, 50\) or \(100\), \(T\) to be \(15, 20\) or \(40\), and \(\sigma\) to be \(0.5\) or \(1\). We repeat each setting \(100\) times to obtain the average performance.

\textbf{Example 2: }In this example, we increase the number of true clusters from \(3\) to \(4\), with the proportions of individuals in the four groups given by \((0.3, 0.2, 0.3, 0.2)\). For each group-specific coefficient vector, the first three entries are set to be non-zero. Different from Example 1, in each iteration, we randomly generate the non-zero parameters as follows. We write the cluster-specific coefficient vectors as \(\bm{\alpha}_{G_k} = (\alpha_{G_k,1}, \alpha_{G_k,2}, \alpha_{G_k,3}, \bm{0}_{p-3})^{\top}_p, k = 1, 2, 3, 4\). For each non-zero coefficient \(\alpha_{G_k,j}\), \(k= 1, 2, 3, 4, j = 1, 2, 3\), we generate it as the product of two random variables. Specifically, we generate \(b_{kj,1} \sim U(0.4,2)\), and independently generate \(b_{kj,2}\) from \(\{-1,1\}\) with each value having probability 0.5. Then we set \(\alpha_{G_k,j} = b_{kj,1}\cdot b_{kj,2}\). We then check whether the \(\ell_{2}\)-distance between every pair of coefficient vectors from the four groups is greater than \(1\), so that the group-specific coefficient vectors are separated. Since only the first three entries are non-zero, this condition also implies that the \(\ell_{\infty}\)-distance between every pair of coefficient vectors from the four groups is greater than \(\frac{\sqrt{3}}{3}\), ensuring separation in terms of the \(\ell_{\infty}\)-distance as well. If this condition is satisfied, we retain the generated coefficient vectors; otherwise, we regenerate them. The generation of the design matrix, the error terms, and the response variable is the same as in Example 1. The choices of \(p, N, T\), and \(\sigma\) are also the same as those in Example 1.
\subsection{Criteria}
Before presenting our simulation results, we introduce the criteria we used in our simulations.

Normalized Mutual Information (NMI) is a criterion used to evaluate the clustering performance. Let \(\mathcal{D} = \{D_1, D_2, \ldots\}\) and \(\mathcal{F} = \{F_1, F_2, \ldots\}\) denote two partitions of all individuals \(\{1,\ldots,N\}\). Then, NMI can be computed by
\begin{gather}
    I(\mathcal{D},\mathcal{F}) = \sum_{i,j}(\vert D_i \cap F_j\vert/N)\log(N\vert D_i \cap F_j\vert/\vert D_i\vert\vert F_j\vert),\nonumber\\
    H(\mathcal{D}) = -\sum_{i}(\vert D_i\vert/N)\log(\vert D_i\vert/N),\nonumber\\
    H(\mathcal{F}) = -\sum_{j}(\vert F_j\vert/N)\log(\vert F_j\vert/N),\nonumber\\
    {\rm NMI}(\mathcal{D},\mathcal{F}) = \frac{2I(\mathcal{D},\mathcal{F})}{H(\mathcal{D}) + H(\mathcal{F})}.\nonumber
\end{gather}
The range of NMI is between \(0\) and \(1\). The closer the NMI is to 1, the better the clustering performance is.

To check the accuracy of coefficient estimation, we use the Estimation Error (EE) of the coefficient vectors (\cite{wang2024homogeneity}), which can be computed by
\begin{gather}
    {\rm EE} = \sqrt{\frac{1}{Np}\sum_{i=1}^{N}\|\bm{\beta}_i - \hat{\bm{\beta}_i}\|^2_2}.\nonumber
\end{gather}

In each replication, we record the number of clusters, denoted by \(K\), as well as the computational time in seconds. The detailed simulation results for \(K\) and computational time are presented in the Supplementary Material.
\subsection{Simulation Results}
In this section, we present the simulation results for Example 1 and Example 2. The boxplots for the setting with \(p = 10\) and \(\sigma = 1\) are shown in Figures 1-4. The boxplots for the remaining settings are provided in the Supplementary Material. The average simulation results are also reported in the Supplementary Material; see Tables A.1 - A.12.

To simplify the notation, in the simulation results below, K-SCAD is the K-regression algorithm with the SCAD penalty. The version of K-SCAD used in our simulations is Algorithm A.2 (Variable Neighborhood Search) proposed by \cite{huang2023integrative}, which the authors describe as being more efficient. The thresholded DPLS algorithm proposed by \cite{wang2024homogeneity} is denoted by tDPLS. Oracle denotes the oracle estimator.

Our simulations are conducted on the High-Performance Computing cluster at the University of Manchester with an Intel Xeon Gold 6130 CPU at 2.10 GHz and 192 GB of RAM, using 25 cores. 
\subsubsection{Example 1}
\begin{figure}
    \centering
    \includegraphics[width=0.8\linewidth]{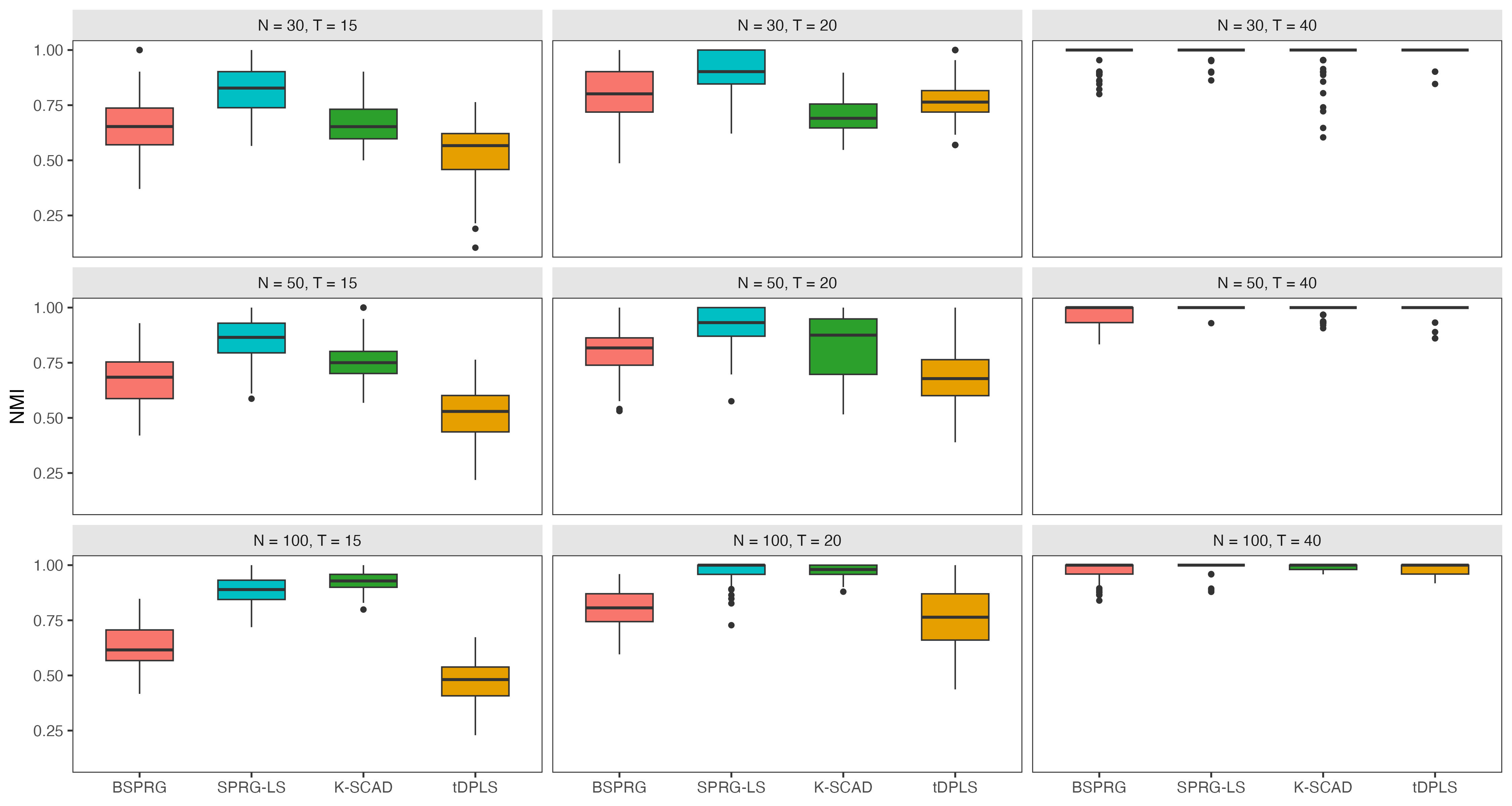}
    \caption{Boxplots of NMI for Example 1 with \(p = 10\) and \(\sigma = 1\), based on \(100\) simulation replications. Red, blue, green, and yellow correspond to BSPRG, SPRG-LS, K-SCAD, and tDPLS, respectively.}
    \label{Example 1 NMI Simulation Results (p = 10 and sigma = 1)}
\end{figure}
\begin{figure}
    \centering
    \includegraphics[width=0.8\linewidth]{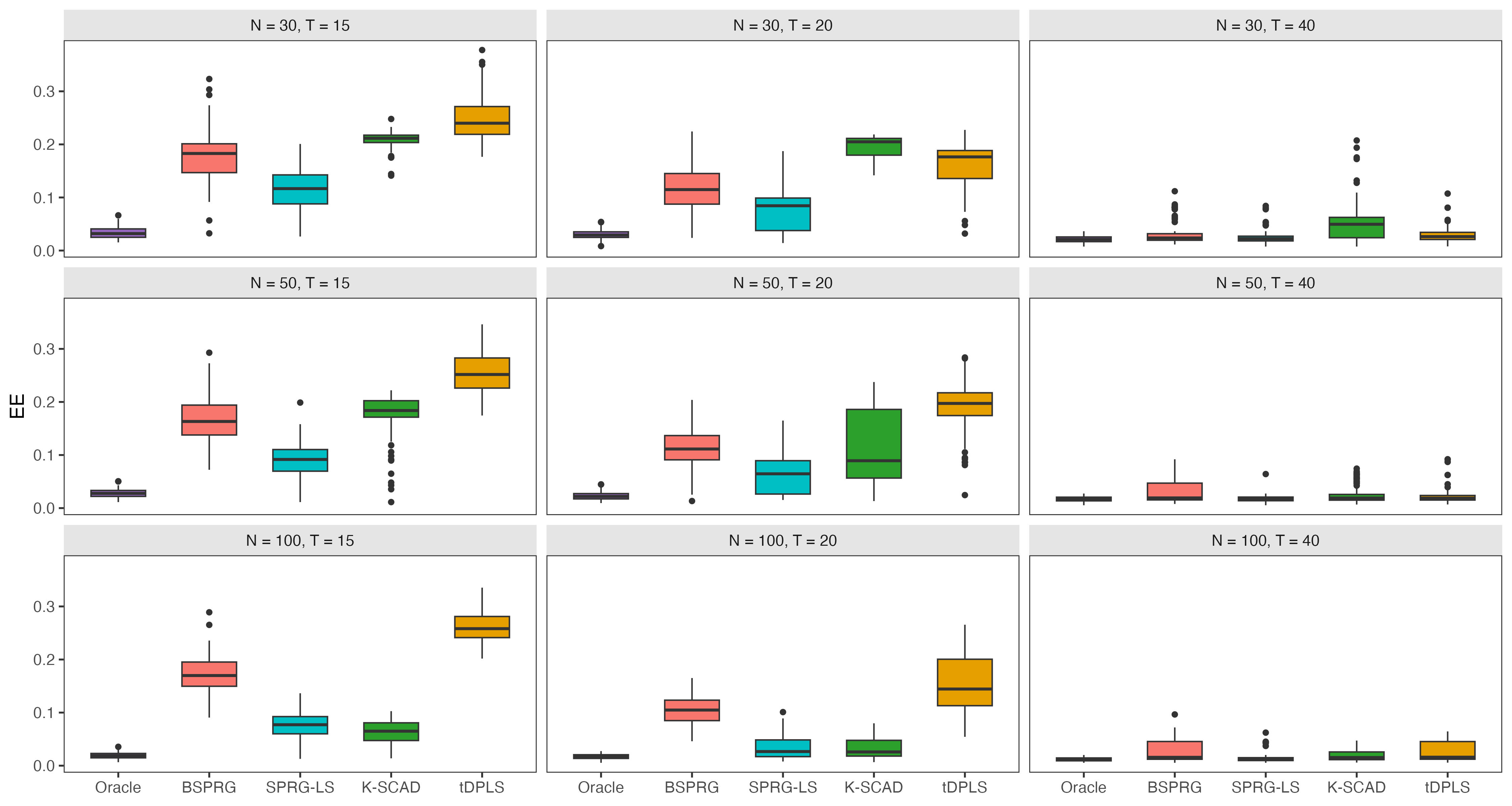}
    \caption{Boxplots of EE for Example 1  with \(p = 10\) and \(\sigma = 1\), based on \(100\) simulation replications. Purple, red, blue, green, and yellow correspond to Oracle, BSPRG, SPRG-LS, K-SCAD, and tDPLS, respectively.}
    \label{Example 1 EE Simulation Results (p = 10 and sigma = 1)}
\end{figure}
When \(p = 10\) and \(\sigma = 1\), \hyperref[Example 1 NMI Simulation Results (p = 10 and sigma = 1)]{Figure \ref*{Example 1 NMI Simulation Results (p = 10 and sigma = 1)}} and \hyperref[Example 1 EE Simulation Results (p = 10 and sigma = 1)]{Figure \ref*{Example 1 EE Simulation Results (p = 10 and sigma = 1)}} show that SPRG-LS has a clear advantage over the other three approaches. When \(T\) is relatively small \((T = 15,20)\), SPRG-LS achieves the highest NMI values and the lowest EE values in most settings. When \(T\) increases to \(40\), all four approaches perform well, with NMI values close to \(1\) and EE values close to those of the oracle estimator. K-SCAD performs second best after SPRG-LS, and in a few cases, its performance is close to or better than that of SPRG-LS, such as when \(N = 100\) and \(T = 15\). BSPRG and tDPLS have relatively similar performance, but both lag behind SPRG-LS and K-SCAD in most cases. When \(\sigma = 0.5\), all four approaches perform well, with NMI close to \(1\) and EE values close to those of the oracle estimator. The detailed results for \(\sigma = 0.5\) are provided in Figure A.1 and Figure A.2 in the Supplementary Material.

When \(p\) increases to \(20\), the performance of all four methods clearly decreases under the same settings of \(N, T, \sigma\). This is reasonable because, as the covariate dimension increases, the estimation accuracy of the coefficients decreases under the same sample size, which leads to worse performance of the methods. Comparing the four methods, SPRG-LS still performs the best, with the highest NMI values and the lowest EE values in most settings. K-SCAD performs second best, while BSPRG and tDPLS lag behind these two methods. The detailed results can be found in Figures A.7, A.8, A.11, and A.12 in the Supplementary Material.

Meanwhile, in terms of computational cost, SPRG-LS also shows an advantage while achieving better performance. Detailed results are provided in Figures A.4, A.6, A.10, and A.14 in the Supplementary Material.

In addition, we further consider a high-dimensional setting with \(p = 100\), and present the performance of BSPRG, SPRG-LS, and K-SCAD under this setting. The corresponding boxplots are presented in Figures A.15 - A.22. The conclusion is consistent with that in the low-dimensional settings, and SPRG-LS still performs the best.
\subsubsection{Example 2}
\begin{figure}
    \centering
    \includegraphics[width=0.8\linewidth]{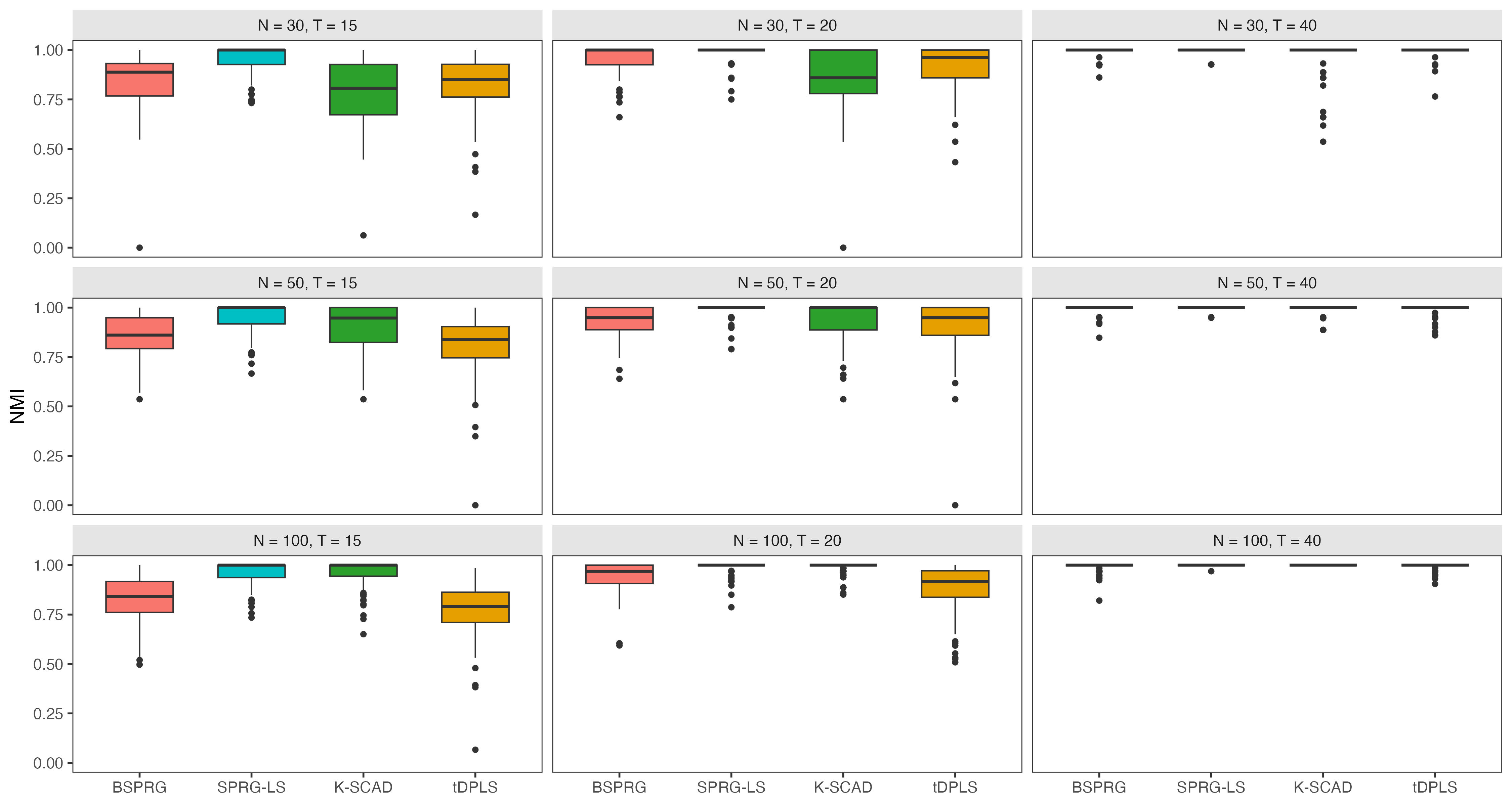}
    \caption{Boxplots of NMI for Example 2 with \(p = 10\) and \(\sigma = 1\), based on \(100\) simulation replications. Red, blue, green, and yellow correspond to BSPRG, SPRG-LS, K-SCAD, and tDPLS, respectively.}
    \label{Example 2 NMI Simulation Results (p = 10 and sigma = 1)}
\end{figure}
\begin{figure}
    \centering
    \includegraphics[width=0.8\linewidth]{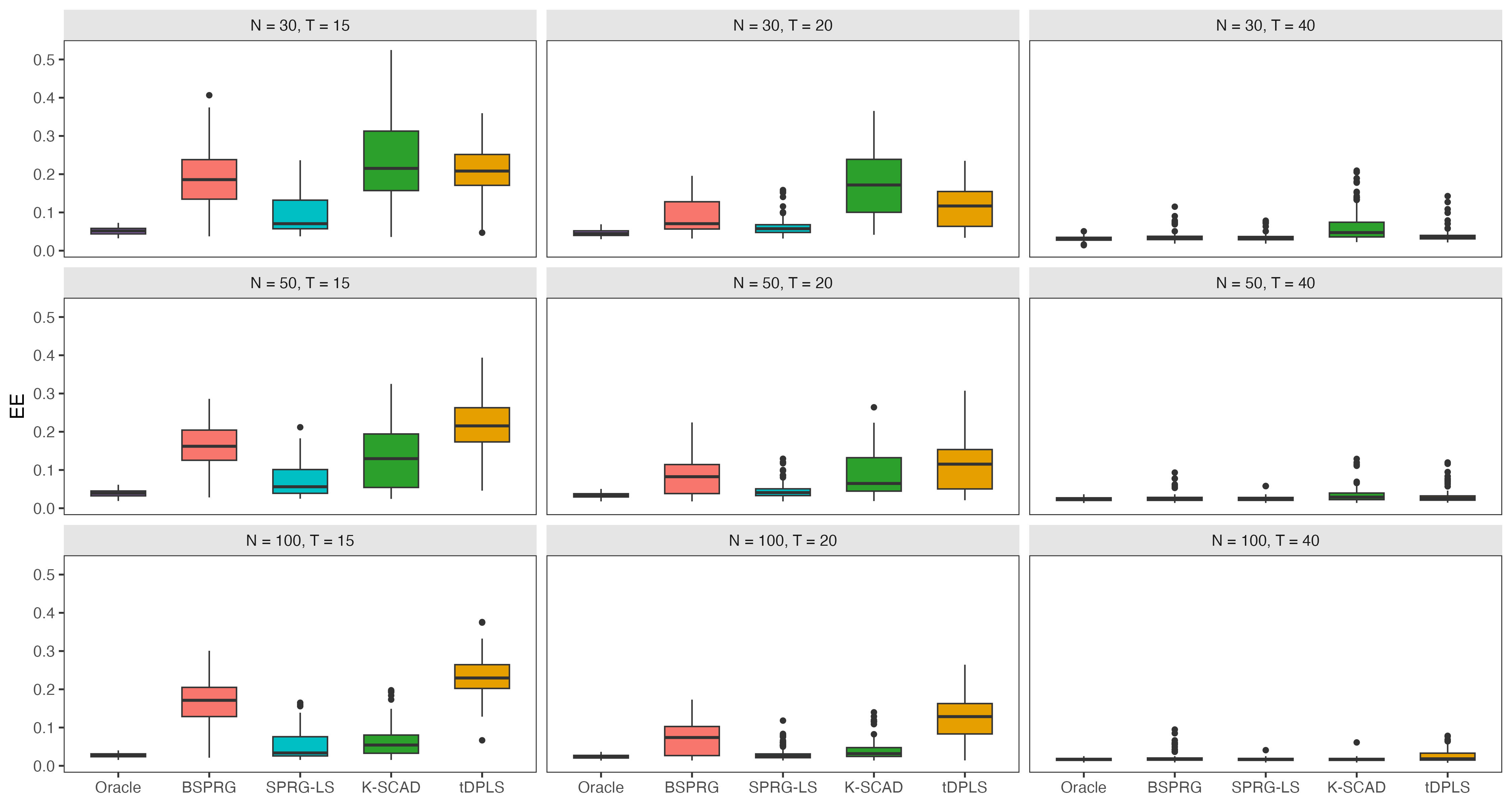}
    \caption{Boxplots of EE for Example 2 with \(p = 10\) and \(\sigma = 1\), based on \(100\) simulation replications. Purple, red, blue, green, and yellow correspond to Oracle, BSPRG, SPRG-LS, K-SCAD, and tDPLS, respectively.}
    \label{Example 2 EE Simulation Results (p = 10 and sigma = 1)}
\end{figure}
Consistent with the results in Example 1, from \hyperref[Example 2 NMI Simulation Results (p = 10 and sigma = 1)]{Figure \ref*{Example 2 NMI Simulation Results (p = 10 and sigma = 1)}} and \hyperref[Example 2 EE Simulation Results (p = 10 and sigma = 1)]{Figure \ref*{Example 2 EE Simulation Results (p = 10 and sigma = 1)}}, we find that when \(p = 10\) and \(\sigma = 1\), SPRG-LS performs better than the other three approaches in most cases, with higher NMI values and lower EE values, while requiring much less computational time. In terms of NMI and EE, K-SCAD performs second best after SPRG-LS, but its computational cost is much higher than that of SPRG-LS. BSPRG and tDPLS perform worse than SPRG-LS and K-SCAD. All four approaches perform well when \(p = 10\) and \(\sigma = 0.5\), and the corresponding boxplots are presented in Figures A.23 and A.24 in the Supplementary Material. As observed in Example 1, when \(p\) increases to \(20\), the performance of all four approaches decreases. Considering NMI, EE, and computational cost jointly, we find that SPRG-LS still performs best among the four approaches. Detailed boxplots can be found in Figures A.29 - A.36 in the Supplementary Material.

Similarly to Example 1, we additionally consider a high-dimensional setting with \(p = 100\). Detailed boxplots are provided in Figures A.37 - A.44 in the Supplementary Material, and SPRG-LS still performs the best.
\section{Real Data Analysis}\label{Real Data Analysis}
In this section, we apply our approach to real-world economic data. Economic growth is influenced by multiple factors. We aim to investigate whether there is a latent group structure among European countries that share similar economic trajectories, while identifying the covariates that truly contribute to economic change. Here, we use the economic data from Penn World Table 11.0 (\cite{feenstra2015next}), which provides comparable measures of GDP per capita across countries. We select a European dataset from Penn World Table 11.0 that covers the period from 2011 to 2020 and includes \(37\) countries and \(20\) covariates. Details about the \(20\) covariates can be found in Table A.13 in the Supplementary Material. We choose the log-transformed GDP per capita as the response. Before fitting the model, we standardize both the response and the covariates.

\begin{figure}[H]
    \centering
    \includegraphics[width=0.8\linewidth]{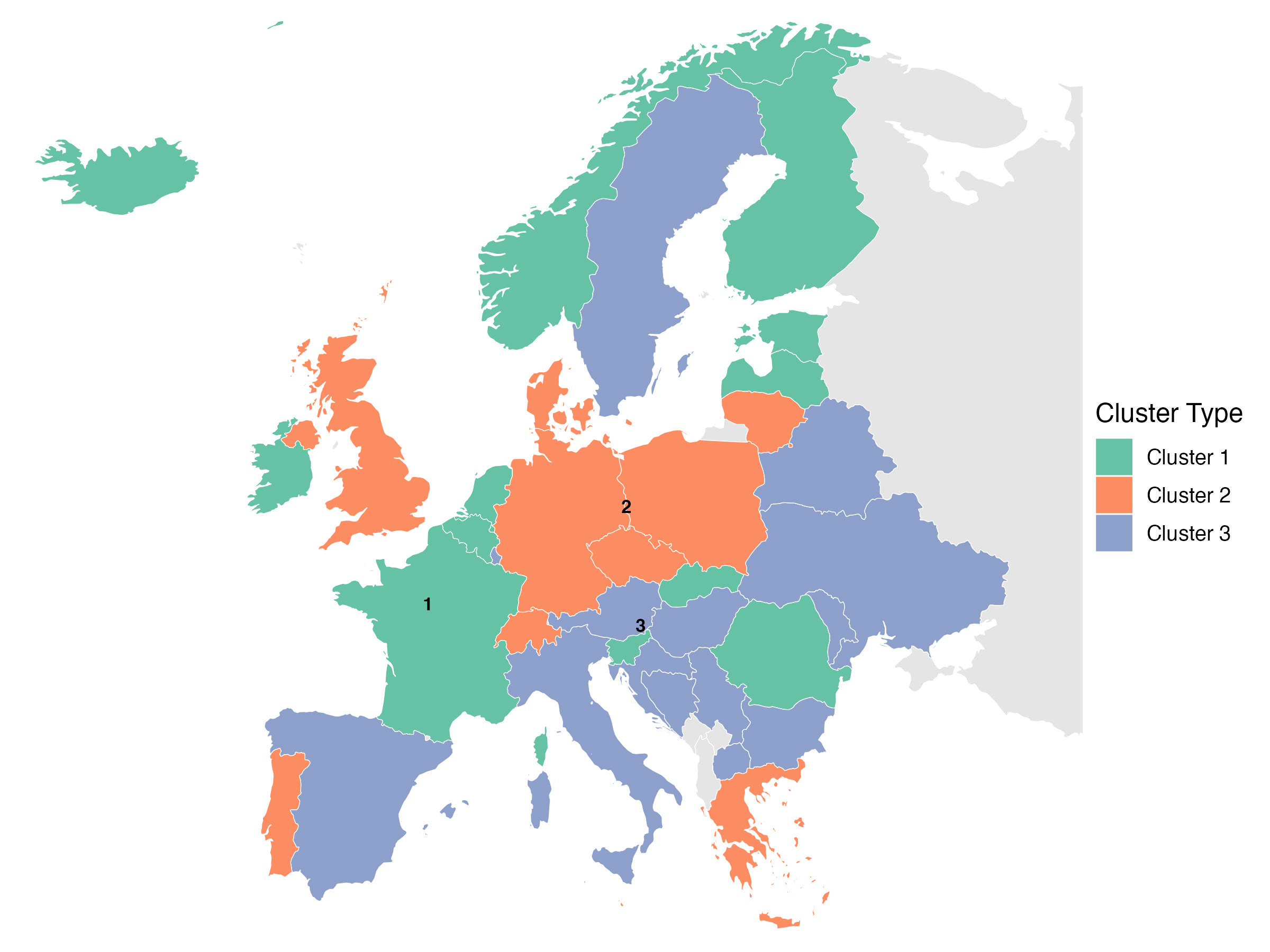}
    \caption{Map of the Clustering Results}
    \label{Real data plot 1}
\end{figure}

After applying SPRG-LS to this dataset, the resulting partition consists of three groups, containing \(12, 11,\) and \(14\) countries, respectively. The detailed cluster results can be found in Table A.14 in the Supplementary Material. Also, to better visualize the clustering results, we present \hyperref[Real data plot 1]{Figure \ref*{Real data plot 1}}. The map is produced using the \texttt{ggplot2} package based on \texttt{sf} spatial objects, with country boundaries obtained from the \texttt{rnaturalearth} package in \textsf{R}.
Based on \hyperref[Real data plot 1]{Figure \ref*{Real data plot 1}}, the identified grouping appears to exhibit a geographical pattern, especially for Clusters 2 and 3. Cluster 1 mainly consists of countries located in Western and Northern Europe, while also including some Central and Eastern European countries, such as Slovakia and Romania. The majority of countries in Cluster 2 are located in Central Europe, while the cluster also includes countries in Western and Southern Europe, such as the United Kingdom, Portugal, and Greece. Cluster 3 comprises many Eastern and Southern European countries, while it also includes Sweden, which is located in Northern Europe.

Next, we present the estimated coefficient vectors for the three clusters. Following \cite{wang2024homogeneity}, we compute the standard errors of the group-specific coefficients using the clustered covariance matrix (CCM) estimator. For details of CCM, see \cite{Arellano1987} and \cite{Hansen2007}.
\begin{table}[H]
    \footnotesize
    \centering
    \begin{tabular}{cccc}
    \hline
    Covariate & Cluster 1 & Cluster 2 & Cluster 3\\
    \hline
    csh\_c & -0.4598 (0.0182) & -                & -0.6334 (0.0348)  \\
    csh\_g & -                & -                & -0.3057 (0.0276) \\
    pl\_g  & 0.4808 (0.0395)  & 0.2610 (0.0459)  & -                \\
    pl\_c  & -                & -                & 0.5022 (0.0313)  \\
    cda    & -                & 0.0112 (0.0388)  & -                \\
    empr   & -                & 0.5543 (0.0642)  & -                \\
    \hline
    \end{tabular}
    \caption{Estimated Group Coefficients}
    \label{Real data table}
\end{table}
Overall, \(6\) covariates are selected by SPRG-LS, and the estimated coefficient vectors for the three clusters contain \(2\), \(3\), and \(3\) selected covariates, respectively. The abbreviations used in \hyperref[Real data table]{Table \ref*{Real data table}} are consistent with those in \cite{feenstra2015next}, and detailed definitions of each abbreviation can also be found in Table A.13 in the Supplementary Material. From \hyperref[Real data table]{Table~\ref*{Real data table}}, we find that \texttt{csh\_c}, which represents the share of household consumption, is selected in both Clusters 1 and 3, with negative estimated coefficients in both groups. This suggests that, for countries in these two clusters, increases in the share of household consumption correspond to decreases in GDP per capita. Meanwhile, \texttt{pl\_g}, which represents the price level of government consumption, has positive estimated coefficients in both Clusters 1 and 2. This indicates that, for countries in these two clusters, a higher price level of government consumption corresponds to a higher GDP per capita. 

The remaining covariates are cluster-specific, and all of them are significant except for \texttt{cda}, which represents real domestic absorption and is not significant in Cluster 2.

\section{Conclusion}\label{Conclusion}
In this paper, we propose a computationally efficient framework for detecting homogeneity and sparsity in coefficient vectors of panel data models. In both BSPRG and SPRG-LS, we make full use of the grouping information contained in the initial estimators obtained from SCAD-penalized regression, with the aim of reducing the computational cost needed to identify homogeneity and sparsity. In addition, SPRG-LS includes a local search correction step to reduce excessive dependence on initial estimators. We provide theoretical guarantees showing that, under reasonable assumptions, our approaches enjoy favourable asymptotic properties. Our simulation studies show that, compared with K-SCAD and tDPLS, SPRG-LS achieves better performance in most settings with shorter computational time. We also apply SPRG-LS to real-world economic data to illustrate its practical applicability. Possible future work includes further relaxing the theoretical restrictions on the initial estimators and on the relationship between the dimension and the sample size, as well as extending our methods to more complex settings, such as nonlinear models and quantile regression models. 

{\large\bf Data Availability Statement}

The data used in \hyperref[Real Data Analysis]{Section \ref*{Real Data Analysis}} are publicly available at \url{https://www.rug.nl/ggdc/productivity/pwt/}.

\phantomsection\label{supplementary-material}
\bigskip

{\large\bf Supplementary Material}

The Supplementary Material contains (i) detailed descriptions of our proposed algorithms, (ii) additional simulation results, (iii) proofs of the asymptotic properties, and (iv) additional information for the empirical application. 

\bibliography{bibliography.bib}
\end{document}